\documentclass[aip,pop,reprint]{revtex4-1}
\usepackage{graphicx}
\usepackage{amsmath}
\usepackage[breaklinks=true,colorlinks=true,linkcolor=blue,urlcolor=blue,citecolor=blue]{hyperref}

\newsavebox{\myimage}
\newcommand\redsout{\bgroup\markoverwith{\textcolor{red}{\rule[0.5ex]{2pt}{0.4pt}}}\ULon}

\usepackage{graphicx}
\usepackage{color}

\begin{document}


\title{The role of incidence angle in the laser ablation of planar ICF targets}


\author{Brett Scheiner}
\affiliation{Los Alamos National Laboratory, Los Alamos, NM 87545}

\author{Mark Schmitt}
\affiliation{Los Alamos National Laboratory, Los Alamos, NM 87545}

\date{\today}

\begin{abstract}

The effect of the laser ray incidence angle on the mass ablation rate and ablation pressure of planar inertial confinement fusion (ICF) targets is explored using an idealized model. 
Polar direct drive (PDD) on the National Ignition Facility (NIF) requires the repointing of its 192 beams clustered within 50 degrees of the poles to minimize the imparted polar varying payload kinetic energy of the target. Due to this repointing, non-normal incidence angles of the beam centerlines are encountered in any PDD design. The formulation of a PDD scheme that minimizes non-uniformity is a significant challenge that requires an understanding of the induced differences in ablation including those of incidence angle. In this work, a modified version of the textbook model of laser ablation [Manheimer et al. Phys. Fluids 25, 1644 (1982)] is used to demonstrate that the mass ablation rate and ablation pressure scale with the 4/3 and 2/3 power of the cosine of the laser ray incidence angle for the planar case during the beginning portion of the laser ablation. This result is considered with an idealized 1-D two segment planar model using rays at different incidence angles for each segment. Within the model, the segments cannot have equal mass and velocity simultaneously without tailoring the ray's time-dependent intensity for each segment. However, with the correct segment-to-segment time-independent ray intensities, equal velocity and dynamic pressure can be achieved approximately without tailoring. Additionally, an analytic prediction for the conduction zone width as a function of incidence angle is provided. It is predicted that this width increases with incidence angle, resulting in a decrease in laser imprint, an effect which has been previously observed in experiments. These results, when generalized to spherical geometry, may provide insight into the implosion dynamics encountered in PDD.

\end{abstract}

\maketitle

\section{Introduction}
\begin{figure}
\includegraphics[scale=.35]{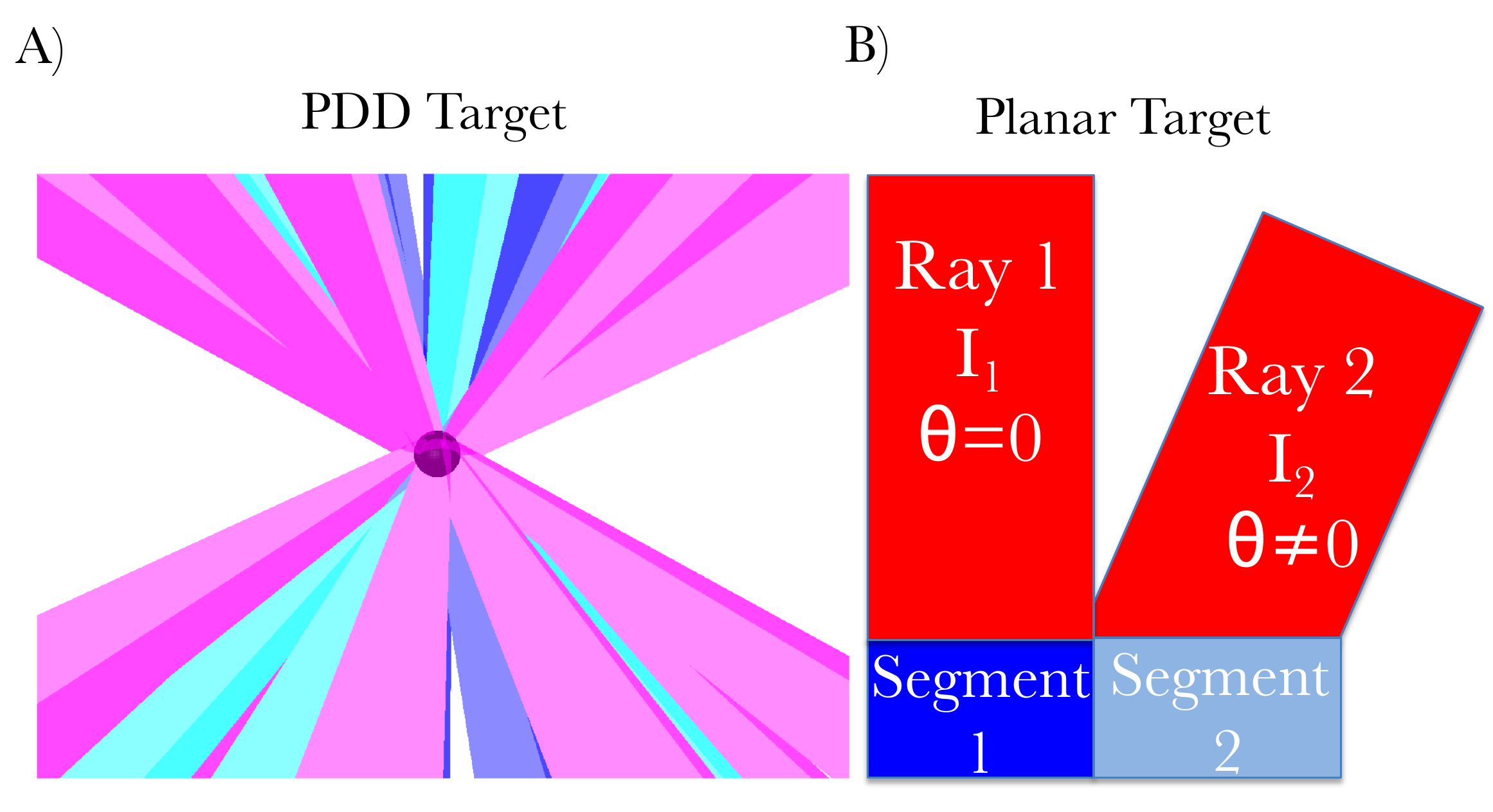}
\caption{A) An OMEGA PDD design modeled with VISRAD\cite{2003JQSRT..81..287M}. The blue, cyan, and purple beams originate from port angles of 21.4, 42.0, and 58.8 degrees from the pole. B) The idealized segmented planar system from this study.\label{pddplane}}
\end{figure}

Polar direct drive (PDD) is the only near-term viable method for driving a variety of proposed direct drive ignition target designs (see Refs. \onlinecite{PhysRevLett.98.155001,PhysRevLett.86.436,PhysRevLett.116.255003,PhysRevLett.119.195001}) on the National Ignition Facility (NIF). In its current indirect drive configuration, NIF's 192 laser beams emanate from laser ports at angles of 23.5, 30, 44.5, and 50 degrees from the poles of its spherical target chamber. In PDD, the pointing of the beam centerlines or beam axes\footnote{Here, the beam centerline is defined as the line perpendicular to the centroid of the beam spot.} on the target surface are redistributed towards the equator in an attempt to provide energy to the target's payload mass with minimal variation in polar angle\cite{2005PhPl...12e6304C,2015PhPl...22e6308H}. An example of a PDD beam pointing scheme is depicted in Fig.~\ref{pddplane}A\footnote{ The beams in this figure utilized a SG-2-600 phase plate to make the repointing easily discernible. Typical experiments on Omega use SG-5 phase plates resulting in larger beam spots.}. The repointed beams ablate a fraction of the target shell thickness resulting in a rocket-effect-like inward motion that compresses the interior fuel, eventually converting the dynamic pressure of the shell into internal pressure of the stagnated fuel. Designing an implosion which avoids imparting asymmetry due to the fixed distribution of beam polar angles is a nontrivial task and has been the subject of experimental testing at both the Omega laser facility and NIF\cite{2016PhPl...23a2711M,2015PhPl...22i2707M} and of many computational studies\cite{2004PhPl...11.2763S,2012PhPl...19e6308C}. 


In PDD and symmetric direct drive, polar and azimuthal variations in drive conditions due to the projection of the beam spots on the capsule seed Rayleigh-Taylor growth, particularly at spherical harmonic modes synergistic with the number of beam polar cone angles ($\ell$ modes) and the number of beams in each cone ($m$ modes). Nonuniformity seeded by low frequency variations in drive conditions is not attenuated to the same extent as high frequency variations. Theoretical models suggest that the attenuation factor is proportional to $\exp(-x_{cz}k)$, where k is the spatial wavenumber of the induced variation and $x_{cz}$ is the length of the conduction zone\cite{2000PhPl....7.2062G}. As a result, high frequency drive variations are attenuated as soon as the conduction zone forms, typically a few hundred picoseconds\cite{2005PhPl...12d0702S}. The same is not true for low-to-mid $\ell$ spherical harmonic modes. When the induced variations have wavelength greater than the conduction zone length the attenuation becomes less significant. Hence, variations in drive conditions with polar angle can result in low-to-mid mode imprint and growth throughout the implosion. The growth of modes in this range increase the fuel-shell interfacial area and atomic level mix across the interface which results in cooling and a reduction in neutron yield.

 A successful application of PDD will require addressing several technical issues that inhibit the reduction of laser-port-distribution-induced asymmetry in this configuration. Of considerable interest in this genre is cross beam energy transfer (CBET), a laser plasma instability that tends to scatter light away from the equatorial regions of the target. The primary interest in CBET is the reduction of energy available to the target, although a secondary concern is the scattering of light away from the equatorial regions which results in increased polar non-uniformity.
 Mitigating these effects has been a major focus in PDD design\cite{PhysRevLett.120.085001,doi:10.1063/1.5022181}. Another issue for PDD is the larger projection of the beam spot for beams repointed to illuminate the equator. This decreases the intensity that can be provided to the target surface by a single equatorial beam and makes the determination of an optimal beam pattern non-trivial. 

Although the illumination of PDD capsules results in beams with different centerline angles of incidence overlapping at any point on the target surface, it is important to first understand differences in how a single beam deposits its energy. In this paper, the role of the ray incidence angle in the scaling of the ablation pressure, mass ablation rate, and electron conduction zone width is explored for an idealized planar target configuration, ignoring multi-dimension effects (see Fig.~\ref{pddplane}B). For the planar configuration, it is found that the mass ablation rate and ablation pressure responsible for accelerating the two disparate portions of the plane scale differently with incidence angle due to the different plasma conditions at the ray's turning point. As a result, if two portions of a segmented plane with uniform material thickness are driven by rays with different incidence angles, strictly equal velocity and mass remaining cannot be obtained by using time-dependent ray intensities that are a scalar multiple of each other. 
In addition, the conduction zone is also found to scale with incidence angle, becoming wider as the off-normal incidence angle increases and implying a reduction in laser imprint for beams with centerlines at large incidence angles, a result which has been previously observed in experiments\cite{2005PhPl...12d0702S}. These results, when generalized to the laser ablation of a spherical target, may inform the design of PDD beam pointings and laser pulse intensity profiles needed for a spherical implosion that minimizes polar beam distribution effects. 

This paper is organized as follows. In Sec.~\ref{ablmodel} a review of the textbook model for ablation formulated by Manheimer et al.\cite{1982PhFl...25.1644M} is given. Sec.~\ref{val} evaluates the validity of the model's assumptions through the use of radiation hydrodynamic simulations.
 The flux balance of this model is modified in Sec.~\ref{roleince} to include the difference in plasma conditions resulting from the change in density at the laser turning point. Section \ref{valinc} revisits the assumptions made in Sec.~\ref{val}, demonstrating that local energy deposition is a better assumption for oblique incidence. Sec.~\ref{drive} illustrates implications of this modification through the formulation of a toy model for the ablation of a segmented semi-infinite plane by two rays at different incidence angles. The scaling of the conduction zone width with incidence angle is presented in Sec.~\ref{czlen}. A discussion of the implication of the toy model for the implosion of spherical targets is presented in Sec.~\ref{impl} and Sec.~\ref{impr} discusses prior experiments which measured the effects of incidence angle on imprint. Concluding remarks are given in Sec.~\ref{conc}.

\section{Normal Incidence}

The textbook model\cite{atzeni2004physics} for the scaling of ablation pressure, mass ablation rate, and conduction zone length with laser intensity, wavelength, and material properties is due to the steady state planar ablative flow model of Manheimer et al.\cite{1982PhFl...25.1644M} (hereafter referred to as the Manheimer model). This section reviews the formulation of the model and evaluates its validity when applied to the laser ablation of spherical targets. Radiation hydrodynamic simulations are used to indicate the duration of validity of the model as the coronal plasma expands and to evaluate the ablative properties for large spherical targets. It is found that the planar model can be appropriately applied to a 3.1 mm radius capsule, similar to the one encountered in the Revolver target baseline design\cite{PhysRevLett.116.255003}, early in the laser pulse. The recovery of planar ablation scaling early in time is attributed to the length scale of the rarefied coronal plasma being small compared to the target radius.
\subsection{Textbook Planar Ablation Model\label{ablmodel}}
\begin{figure}
\includegraphics[scale=.55]{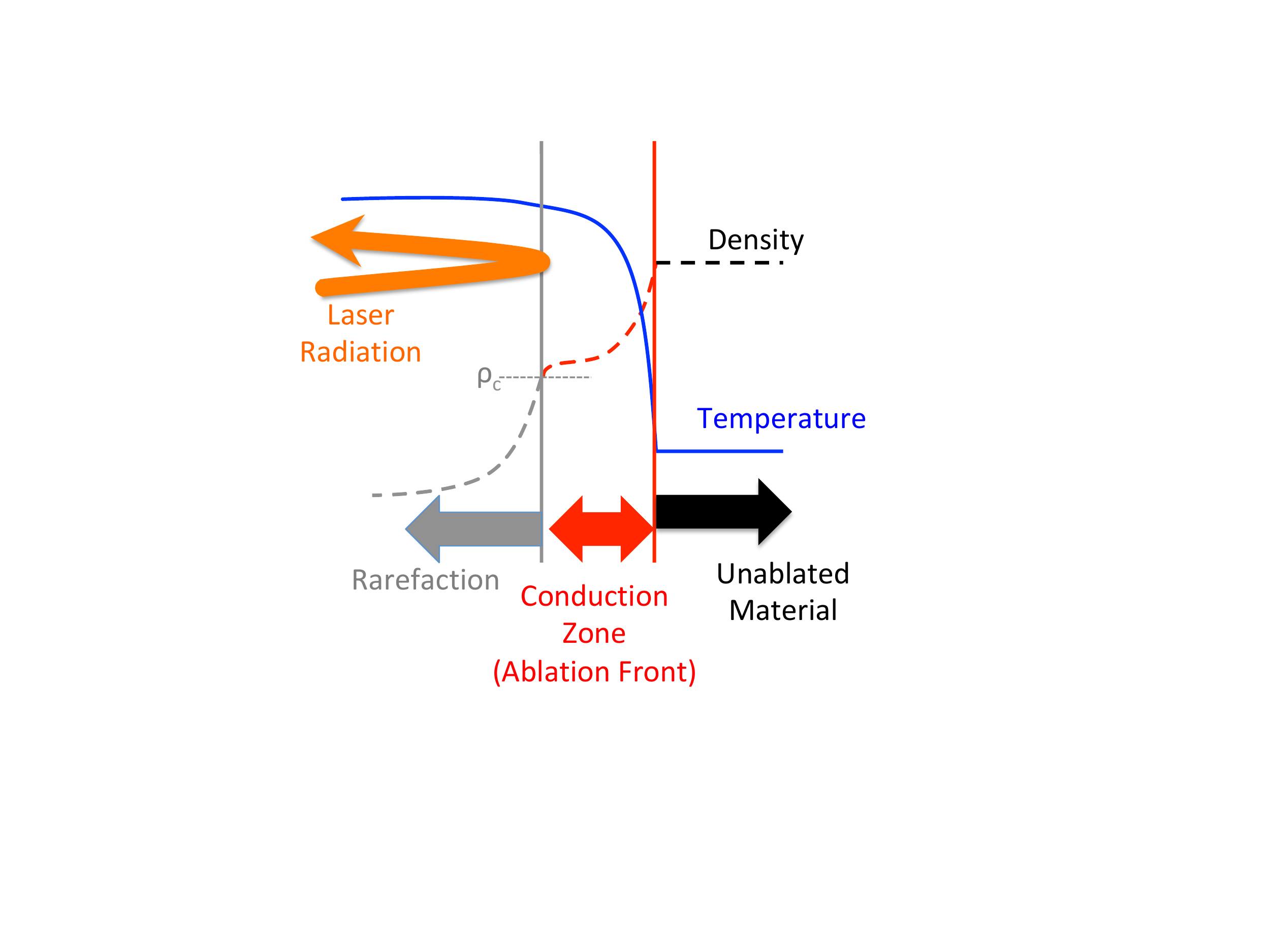}
\caption{An illustration of the electron density and temperature in the planar ablative flow model with the laser turning point at the critical density $\rho_c$. \label{figabl} }
\end{figure}

In the Manheimer model, normally incident laser energy is assumed to be absorbed at the critical surface, the classical turning point of the laser beam rays where the plasma frequency matches the laser radiation frequency. The model specifies a relation between the laser intensity and coronal plasma properties by considering the expected physical behavior inward and outward of the laser turning point, enforcing a power balance relation between these regions. The main features of the model are depicted in Fig.~\ref{figabl} and are described in this section.

In the region outward of the critical density the expanding plasma is expected to be well modeled by a self-similar rarefaction. Since the plasma in this region has a high conductivity, the rarefaction is treated as isothermal and has a density and velocity profile given by the self-similar solution 
\begin{equation}\label{ssrho}
\rho=\rho_c\exp(-x/c_Tt)
\end{equation}
 \begin{equation}
 v=-c_T+x/t,
 \end{equation} where $\rho_c$ is the critical density and $c_T$ is the ion sound speed. 
 
In the region inward of the turning point is the ablation front or conduction zone, which is modeled as a steady state flow\footnote{In Ref. \onlinecite{1982PhFl...25.1644M} the term ablation front is used to refer to the region of steady state flow driven by electron thermal conduction. In the ICF literature, many authors use term ablation front to refer to the ablation surface where the temperature of the target rapidly increases from near zero. }. In the model, any energy conducted inward ends up in the outward flow energy, and therefore in the rarefaction. However, in reality a few percent of the energy conducted inward goes into the kinetic energy of the shell. At the boundary between the steady state region and the rarefaction wave the flow becomes supersonic and the steady state momentum and continuity equation solutions become singular. At this point the power balance is considered, including that of the flow and incoming laser radiation. The energy flux in the flow is given by $(v^2/2+h)\rho v$,  where $h$ is the specific enthalpy. Making use of the condition $v=c_T$ at the turning point, and $h=5c_T/2$, the power being input into the rarefaction from the flow is $P_{flow}=3\rho_cc_T^3$. On the other hand, the total energy in the rarefaction wave can be determined by integrating the kinetic and internal energy over its extent from 0 to $-\infty$. This results in $E_r=4\rho_cc_T^3t$, which implies that a power per unit area of $P_{in}=4\rho_cc_T^3$ is needed to keep the rarefaction wave isothermal. This power is provided by the laser radiation at the deposition point. The power balance condition at the turning point must be 


\begin{equation}\label{A}
I_L=4\rho_c c_T^3,
\end{equation}
where $I_L$ is the laser intensity. This equation is responsible for the well known scalings of ablation pressure 
\begin{equation}\label{pa}
P_a=2\rho_c c_T^2=\bigg(\frac{\rho_c}{2}\bigg)^{1/3}I_L^{2/3},
\end{equation}
mass ablation rate 
\begin{equation}\label{ma}
\dot{m}_a=\rho_cc_T=\bigg(\frac{\rho_c}{2}\bigg)^{2/3}I_L^{1/3},
\end{equation} 
and temperature
\begin{equation}
\Gamma T_c = c_T^2=\bigg(\frac{I_L}{4\rho_c}\bigg)^{2/3},
\end{equation}
where, $\Gamma=(1+Z)k_B/(Am_p)$, A is the number of nucleons per atom in the target material, Z is the ionization state, $\rho_c$ is the critical density, and $T_c$ is the isothermal temperature at the critical density.

The steady state flow inward of the laser turning point is modeled using the steady state momentum and continuity equations along with the energy equation 
\begin{equation}
\frac{d}{dx}\bigg[\bigg(\frac{v^2}{2}+h\bigg)\rho v-\chi_0T^{5/2}\frac{dT}{dx}\bigg]=0,
\end{equation}
where $\chi_0$ is the conductivity. In the energy equation, the Manheimer model neglects the $v^2$ term so that a simple analytic solution can be obtained. This results in a 20 percent error at the critical surface where this term is largest. Enforcing the continuity of mass flux $\rho v=\rho_c(\Gamma T_c)^{1/2}$, the solution for temperature in the ablation front takes the form
\begin{equation}
T(x)=T_c\bigg(1-\frac{25}{4}\frac{\rho_c(\Gamma T_c)^{3/2}}{\chi_0T_c^{7/2}}x\bigg)^{2/5},
\end{equation}
which is valid for $x<x_{cz}$ where
\begin{equation}\label{xaf}
x_{cz}=\frac{4\chi_0}{25\rho_c\Gamma^{3/2}}T^2_c.
\end{equation}
This is the width of the conduction zone. 

\subsection{Validity of the model \label{val}}

The Manheimer model presents a picture where the laser energy has a delta function deposition at the critical surface with no interaction between the incoming laser radiation and the expanding rarefied plasma. In more realistic computational models of ICF experiments this is hardly the case. Although the absorption rate is strongest at the critical surface due to the resonance between the radiation frequency and the plasma frequency, significant laser absorption can occur along the laser's path length well before the critical surface is reached. In particular, the amount of laser absorption away from the critical surface increases the longer the rarefaction is driven. One can easily verify from the WKB solution of laser absorption that this is indeed the case\cite{kruer1988physics}. The increasing value of $c_s t$ in Eq.~\ref{ssrho} means that more plasma near but less than the critical density, with plasma frequency near but less than the resonance condition, is participating in the inverse bremsstrahlung absorption of the laser energy. 

This fact raises the question of whether or not the delta function absorption is a sufficient treatment of the system. Several experiments have attempted to measure the ablation pressure or mass ablation rate either directly or through derived quantities\cite{1983PhFl...26.2011K,1982OptCo..42...55G,1983PlPh...25..237M,doi:10.1063/1.3646554}. Although there is an abundance of experimental results, most of the measurements utilized short pulse times and therefore the results are valid where $c_s t$ is small. Additionally, the intensities that have been explored in these experiments is a few times $10^{13} \textrm{W} \ \textrm{cm}^{-2}$, an order of magnitude lower than that encountered in direct drive ignition concepts. At such low intensities, there is the possibility that the ablation would be best described by strong inverse bremsstrahlung (SIB), which has the ablation pressure scaling of
\begin{equation}\label{sibpa}
P_a\propto I_L^{7/9}
\end{equation}
with absorbed intensity in spherical geometry\cite{1968PlPh...10..867C}. In this section, simulations are employed to determine the range of validity of the Manheimer model of ablation.

\begin{figure}
\includegraphics[scale=.5]{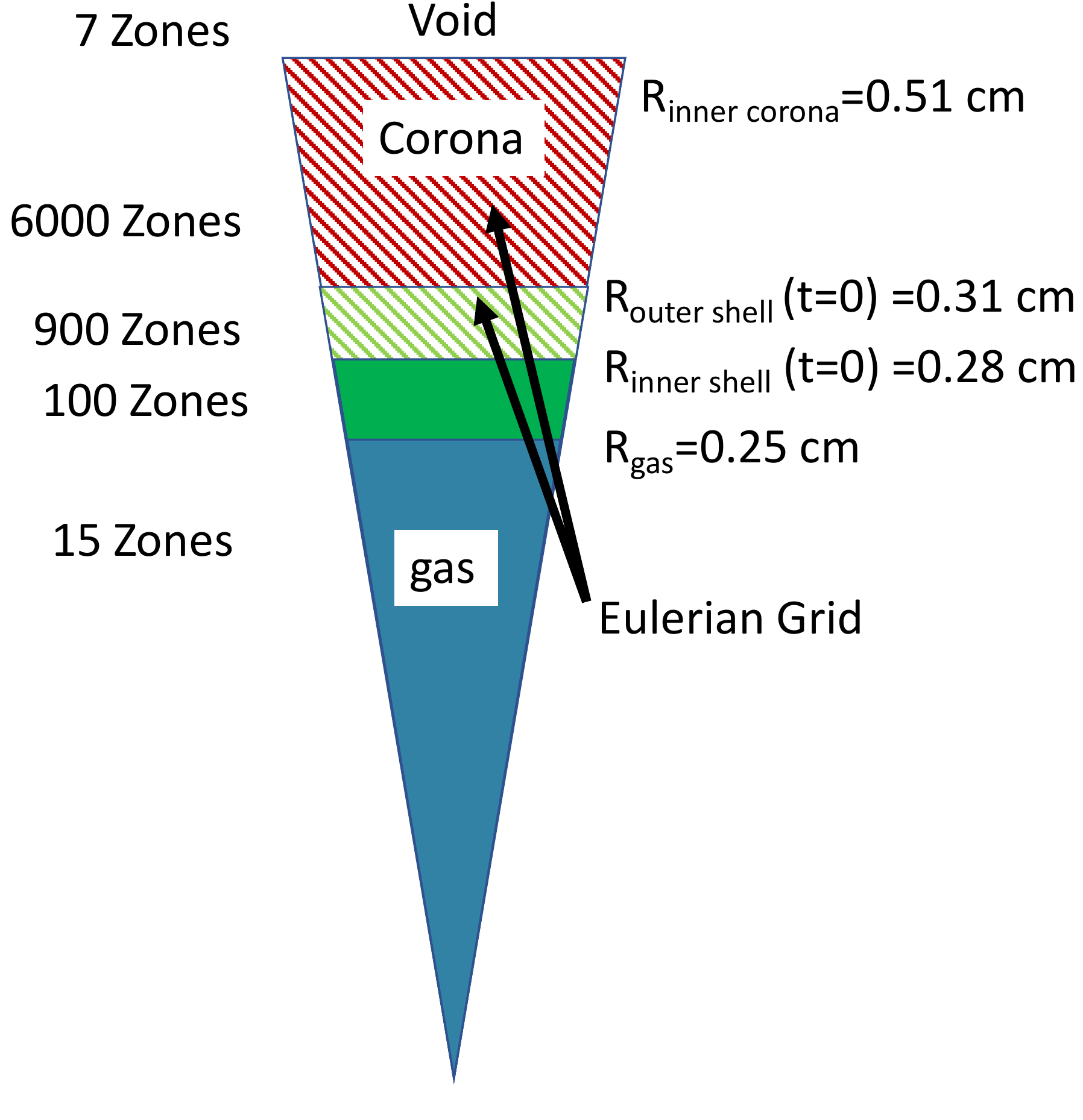}
\caption{The setup of the 1-D spherical coordinate simulation domain indicating the location of the Eulerian grid used to resolve the ablation region. \label{simsetup}}
\end{figure}

Simulations of the normal incidence laser ablation of a 3mm radius 300$\mu m$ thick Be target were carried out using the arbitrary Eulerian Lagrangian radiation hydrodynamic code HYDRA\cite{1996PhPl....3.2070M}. The outer dimension of the target being chosen to resemble the Revolver target design\cite{PhysRevLett.116.255003}$^,$\footnote{The baseline revolver design utilizes a foot to the main pulse, however, some variations omit the foot. } which will be shown to have 1-D ablation characteristics early in the ablation. The primary consideration of the simulation design was to resolve the laser ablation region at all times. This was achieved by considerations in the shell thickness and mesh. First, the target was made sufficiently thick so that little motion of the shell would occur during the laser pulse. The excessive thickness of the shell extended the shock transit time compared to that of typical ICF designs, preventing excessive motion of the shell. In addition, a semi-Eulerian mesh with fixed spacing was implemented in the ablation region as well as the outer portion of the shell by setting the initial zoning to the desired resolution and then backing up the grid nodes to their original position each timestep. The semi-eulerian mesh was required to maintain resolution in the ablation region where Lagrangian zones tend to rapidly expand while conserving mass. The other parts of the problem utilized a Lagrangian mesh. A quasi-1D mesh was implemented in spherical coordinates by simulating a one square degree wedge with appropriate boundary conditions at the wedge faces to ensure the proper symmetry. The mesh dimensions and spacing are given in Fig.~\ref{simsetup}. 

A flat-top laser pulse of duration equal to the length of the simulation was used for each of the simulations presented in this section. The simulations had variable duration, with the extent determined by the required time to demonstrate the desired features discussed below. While many ICF designs utilize a foot or pickets, or combinations thereof, such features are transient and obscure the physics of laser ablation by making time dependence inherent to the system. The case of primary interest was that with intensity characteristic of the Revolver capsule design $I_o=2.65\times10^{14}\textrm{W} \ \textrm{cm}^{-2}$. Simulations were carried out at intensities of $I_o$ as well as multiples of 0.2, 0.4, 0.6, 1.2, 1.4, 1.6, 1.8, 3.0 and 4.0 $I_o$.

\begin{figure}[h!]
\includegraphics[scale=.5]{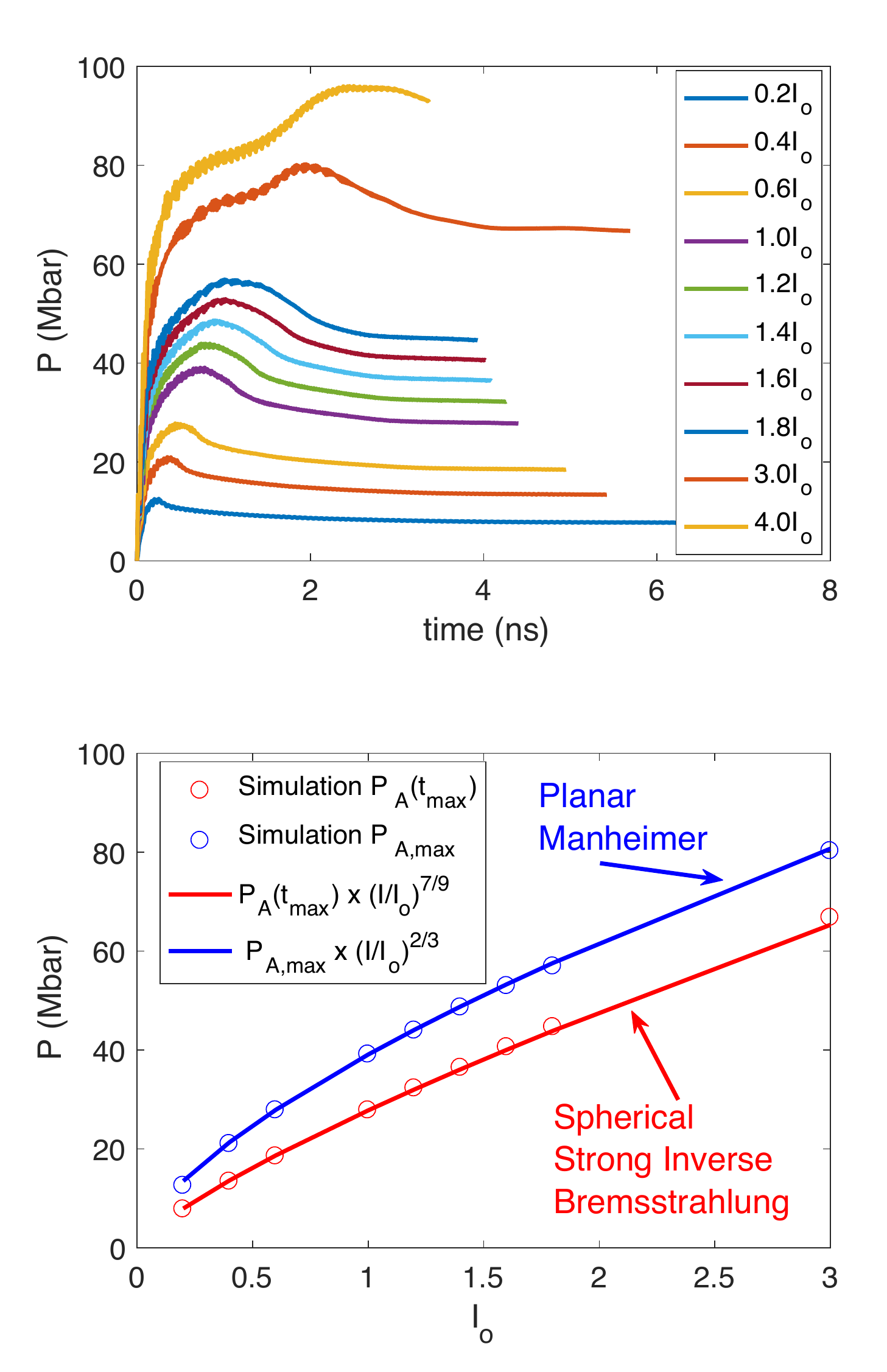}
\caption{Top: The time dependent ablation pressure for multiple intensities. Bottom: The scaling of  $P_{A,max}$ and $P_A(t_{max})$ as a function of intensity for both the Manheimer model and SIB.   \label{pafig}}
\end{figure}

The behavior of the simulations, as characterized by the time dependent ablation pressure, generally falls within either Manheimer or SIB ablation, traversing both regimes for all intensities explored given a long enough pulse length. A distinction between Manheimer and SIB ablation can be made by evaluating the time dependent profiles of the ablation pressure shown in Fig.~\ref{pafig}. As seen in the figure, each profile exhibits an initial peak in ablation pressure before a decay to a nearly constant value which persists until the end of the laser pulse. The constancy of the ablation pressure is a feature of models for Manheimer ablation in a planar expansion as well as SIB ablation for a spherical expansion, however, it will be shown that the former only occurs as a transient state while the latter can occur as a steady state.

To distinguish the two regimes of laser ablation, features in the time dependent ablation pressure profile are identified and plotted as a function of intensity. The first is the peak ablation pressure $\textrm{max}(P_A)$ and the second is the ablation pressure at the end of the simulation run $P_A(t_{max})$. The maximum ablation pressure is a well defined quantity while $P_A(t_{max})$ is not. However, once a steady value of ablation pressure is reached, $P_A(t_{max})$ changes little in time. Fig.~\ref{pafig}B shows the scaling of each  $\textrm{max}(P_A)$ and $P_A(t_{max})$ as a function of laser intensity. The figure also includes the curves representing the quantities 
\begin{equation}
\textrm{max}(P_A(t,I_o))\times \bigg(\frac{I}{I_o}\bigg)^{2/3}
\end{equation}
and
\begin{equation}
P_A(t_{max},I_o)\times \bigg(\frac{I}{I_o}\bigg)^{7/9}
\end{equation}
which indicate that the values of $\textrm{max}(P_A)$ and $P_A(t_{max})$ have the scaling given by Eqs. \ref{pa} and \ref{sibpa}, for Manheimer and SIB ablation, respectively. The agreement with the data indicates that {\emph{planar-like Manheimer ablation is present early in the laser pulse only as a transient phase while SIB ensues given a long enough pulse duration.}} Even then, the Manheimer-like ablation never reaches a steady ablation pressure, even for shorter high intensity pulses similar to those that dominate the parameter space for ignition scale direct drive implosions. Such behavior illustrates the difficulty of using a single intensity scaling law for the entire pulse duration. 

\begin{figure}[h!]
\includegraphics[scale=.5]{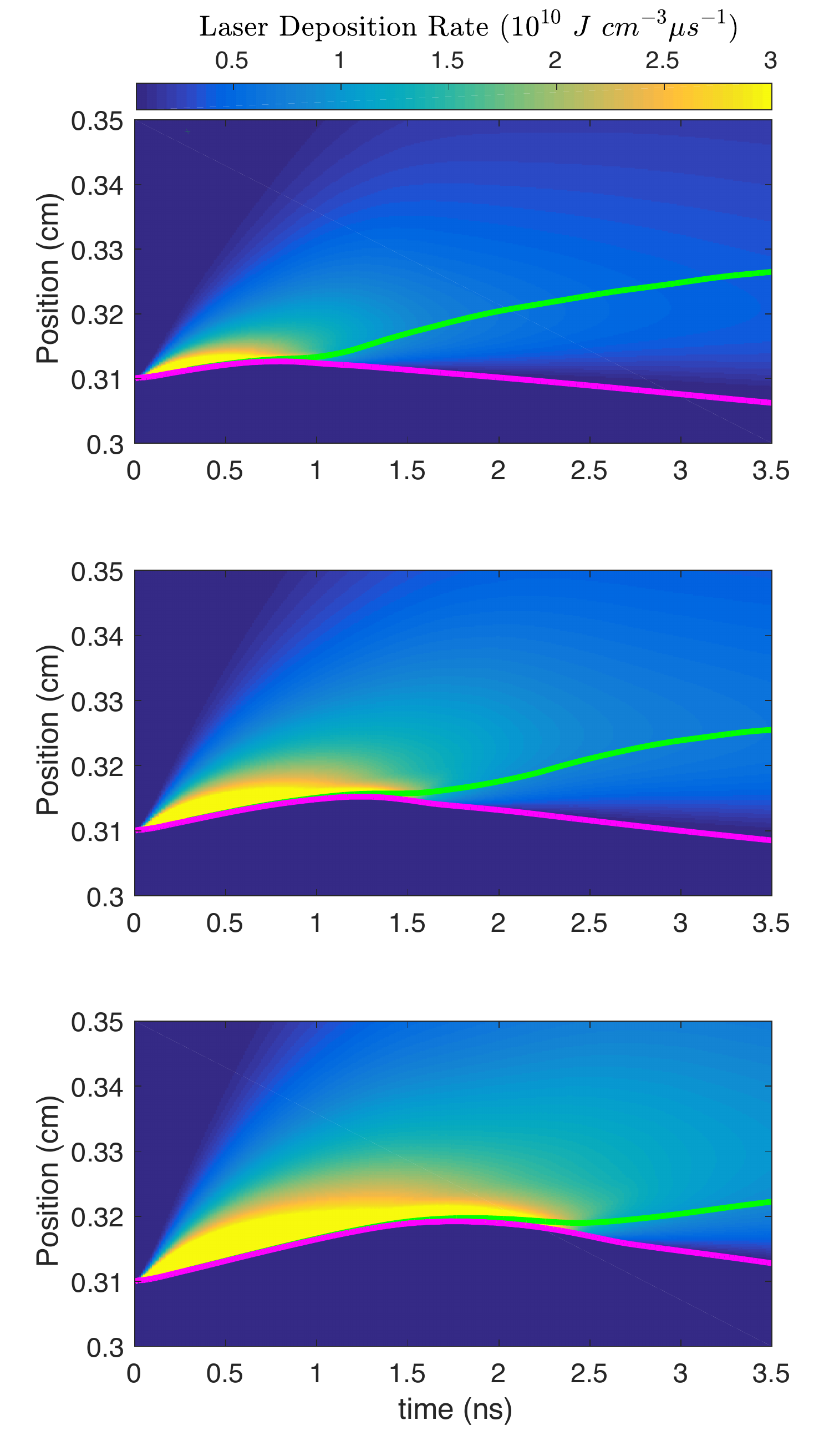}
\caption{The spatial dependence of laser deposition rate in the corona as a function of time for intensities of $I_o$ (top), $1.8I_o$ (middle), and $3.0I_o$ (bottom). The green and purple lines indicate the locations of the sonic point and critical surface, respectively. The deposition becomes distributed across the corona much faster at lower intensity \label{elvplot}}
\end{figure}

It is worthwhile to note that the transition to SIB ablation scaling only occurs once the laser energy deposition rate is no longer strongly peaked near the critical surface as shown in Fig.~\ref{elvplot}. Once a large enough value of $c_st$ is achieved, nearly all of the laser energy is deposited in the plasma volume away from the critical surface (indicated by the purple line in Fig.~\ref{elvplot}). Once the transition occurs, there is no discernible significance of the sonic point (indicated by the green line in Fig.~\ref{elvplot}) in the laser deposition profile. The coincidence of the critical surface and sonic point is a key feature in Manheimer ablation. {\emph{The separation of the location of the sonic point and critical density is therefore a marker of the transition to SIB ablation.}} Although simulations indicate that it is an initial transient phase, it will be assumed that Manheimer ablation is an accurate description of the laser pulse in the initial 1.5 ns for the intensity of $2.65\times10^{14}\textrm{W} \ \textrm{cm}^{-2}$ studied in the following section. Additionally, the result that a large spherical target can reproduce the planar ablation pressure scaling will play a role in the discussion of Sec.~\ref{impl}.

\section{Effect of Incidence Angle On Planar Ablation\label{roleince}}
In PDD on OMEGA or NIF, centerlines of laser beams from either three or four different cone angles intercept the target at a range of incidence angles as shown in Fig.~\ref{pddplane}A. It is also important to note that there is also a second incidence effect for beam spots which are similar in size compared to the target radius. In this case, the laser rays will intercept the target at different incidence angles due to the finite width of each beam and the capsule curvature. However, these effects of curvature are not studied in this paper.

In this section, an idealized planar system is employed to understand the difference in drive conditions for regions illuminated by rays at different incidence angles, see Fig.~\ref{pddplane}B. The model utilized here will be valid early in time when the assumption of localized energy deposition is valid. Understanding drive induced non-uniformity at such times is particularly important since imprint from unattenuated modes will have a longer duration to grow. Furthermore, once the assumption of localized energy deposition is no longer valid, greater attenuation of these modes occurs. 

In the idealized system, the two rays, one normal and one at incidence angle $\theta$, are assumed to independently illuminate two different segments of a plane, with no overlap between the illuminated areas. Ray 1 and ray 2 provide surface intensities $I_1$ and $I_2$ to segments 1 and 2, respectively. The intensity of each ray is assumed to be constant in time. In addition to this setup, the system is considered to be 1D and semi-infinite under each segment, with no cross-conduction smoothing the energy drive between the regions and no distributed inverse bremsstrahlung absorption. Ray 1, which is normally incident on the plane, provides an ablation pressure and mass ablation rate given by Eqs.~\ref{pa} and \ref{ma}. For ray 2 the scaling relation needs to be examined more carefully. 
The turning point for a laser ray with oblique incidence angle is given by $\epsilon(z)-\sin^2\theta=0$, where $\epsilon(z)=1-\omega_{pe}^2(z)/\omega^2$ is the high frequency plasma dielectric assuming a stationary ion background, and $\omega_{pe}(z)$ is the plasma frequency which is spatially dependent on the electron density profile. 
This implies the well known result that the density at the turning point for an obliquely incident ray is $\rho_{tp}=\rho_c\cos^2\theta$.
The significance of the critical density in the Manheimer model is that it represented the density at which all of the laser energy is deposited and, therefore, sets the power balance. 
The dependence of coronal plasma conditions on the ray incidence angle can be easily obtained by writing the turning point density in terms of the critical density for the model in Sec.~\ref{ablmodel}. 
Therefore, the condition of Eq.~\ref{A} can be replaced with one at the oblique ray's turning point, 
\begin{equation}\label{B}
I_L=4\rho_c\cos^2\theta \ c_{T,2}^3,
\end{equation}
where  $c_{T,2}$ is the sound speed based on the temperature at the turning point of the obliquely incident ray 2. Assuming complete absorption of the laser intensity at the rays turning point, the result is that the same amount of ray energy per unit area is deposited into a lower density plasma. See the Appendix for more details on the assumption of complete absorption.

\subsection{Validity of the Manheimer model for oblique incidence \label{valinc}}

The 1-D spherical geometry simulations presented in section \ref{val} demonstrated that the ablation pressure scaling of the Manheimer ablation model given in Eq.~\ref{pa} is realized early in time when the corona length scale is such that the absorption of laser energy occurs near the critical surface. One concern is that at oblique incidence  the absorption of laser radiation will not be as localized as in the normal incidence case. In this section we digress temporarily to demonstrate that for oblique incidence the absorption is {\emph{more}} localized in the direction normal to the target surface than the case of normal incidence. 

The amount of laser absorption due to inverse bremsstrahlung from $\infty$, where the density goes to zero, to some point x in the density profile is given by\footnote{This is a modified verison of Eq. 11.35 of Ref. \onlinecite{atzeni2004physics}. The present equation results from zeroing out the density profile between 0 and x to give the contribution from x to $\infty$. } 
\begin{equation}\label{alx}
A_L(x)=1-\exp\bigg(-2\int_{x}^{\infty} Im\big(k(x)\big)dx\bigg)
\end{equation} 
where $Im(k)$ is the spatial absorption rate given by the imaginary part of the dispersion relation $k(w)$ with $w$ being a function of x through the density profile. The spatial absorption rate for oblique incidence is 
\begin{equation}
Im(k)=\bigg(\frac{\rho}{\rho_{tp}}\bigg)^2\frac{\nu_{ec}\cos^3\theta}{c}\frac{1}{\sqrt{1-\rho/\rho_{tp}}},
\end{equation}
where $\nu_{ec}/c$ is the collision frequency at the critical density given by Eq.~\ref{nuec}. Inserting the density profile $\rho=\rho_{tp}\exp(-x/c_Tt)$, assuming an isothermal rarefaction from the turning point outward, and evaluating the integral in Eq.~\ref{alx} leads to 
\begin{eqnarray}
A_L(x)=1-\exp\bigg(-\frac{4L\cos^3\theta}{3}\frac{\nu_{ec}}{c}\times\nonumber\\
\bigg[ 2-\sqrt{1-e^{-x/L}}(2+e^{-x/L})\bigg]\bigg),
\end{eqnarray}
where $L=c_Tt$ is the plasma scale length. 

\begin{figure}
\includegraphics[scale=.4]{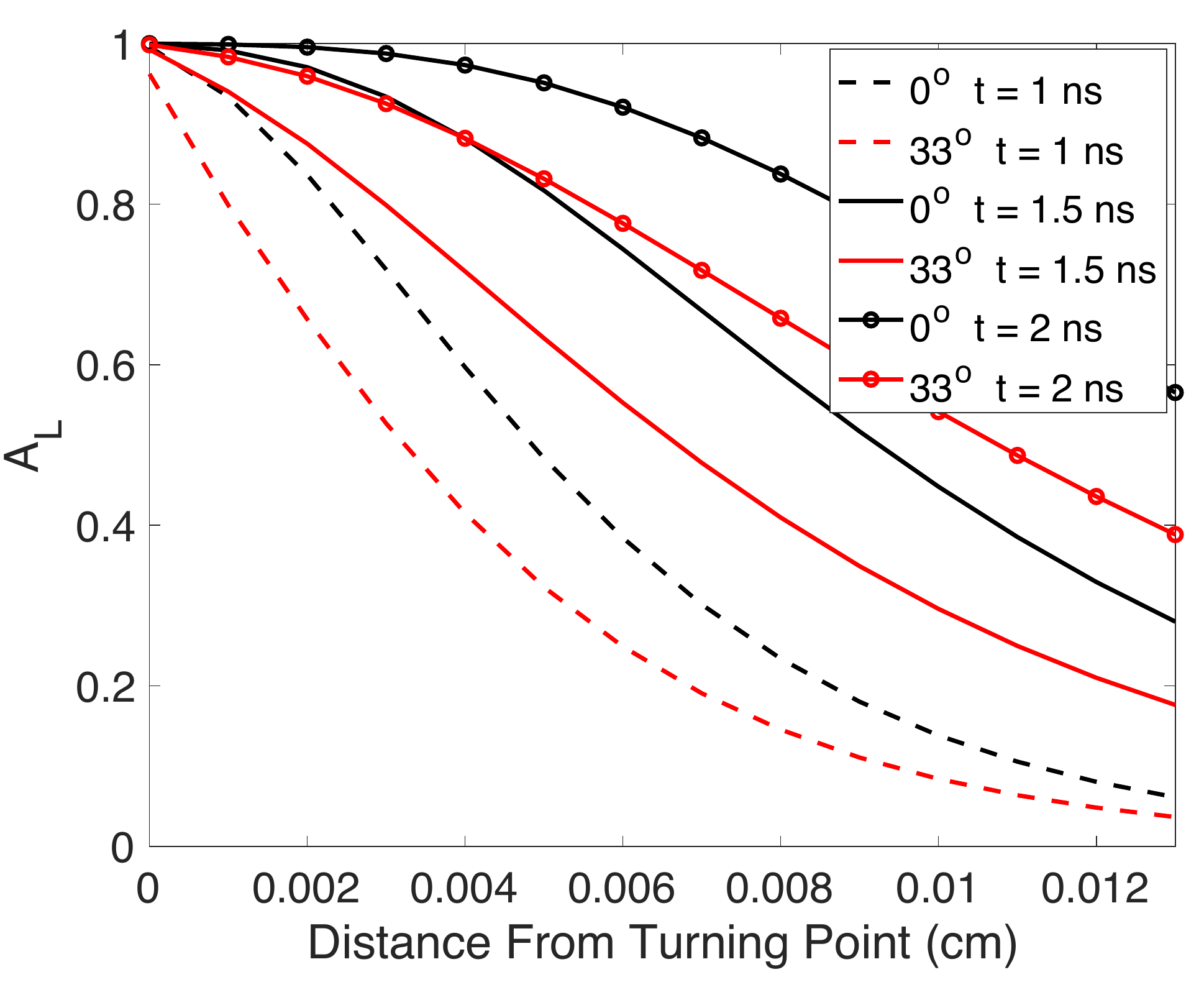}
\caption{The calculated absorbed light fraction for the region of the corona extending from a distance  from the turning point to infinity. Results for times of 1, 1.5, and 2 ns at incidence angles of 0 and 33 degrees. \label{dAlfig}}
\end{figure}

Calculations of the profile of absorption fraction at 1 ns, 1.5 ns, and 2 ns are shown in Fig.~\ref{dAlfig} for both cases of normal and 33 degree incidence in a fully ionized Be plasma, assuming that $T_e=2 keV$. For these parameters, the collision frequency at the critical density is $\nu_{ec}/c=287 \ \textrm{cm}^{-1}$ and the plasma length scale is $L=146\times(t/\textrm{ns}) \ \mu m$. The figure indicates that the majority of absorption is within $50 \ \mu m$ of the turning point for the first nanosecond. {\emph{An even more striking feature of the figure is that the absorption at oblique incidence occurs closer to the ray turning point density than for the normally incident case.}} This effect results from the ray spending more time near its turning point density due to its prolonged propagation in the transverse direction to the density gradient. As a result, localized deposition near the turning point is a better approximation for obliquely incident rays than for normal rays. 

\subsection{Conditions for minimizing drive non-uniformity\label{drive}}
With the modified relation given in Eq.~\ref{B}, the mass ablation rate and ablation pressure for oblique incidence can be expressed as
\begin{equation}
\dot{m}_a=0.61\frac{\textrm{g}}{\textrm{cm}^2\mu\textrm{s}}\bigg(\frac{A\cos^2\theta}{2Z}\frac{n_c}{n_{351}}\bigg)^{2/3}\bigg(\frac{I_L}{I_{14}}\bigg)^{1/3}
\end{equation}
and
\begin{equation}
P_a=24.7\textrm{Mbar}\bigg(\frac{A}{2Z}\frac{n_c}{n_{351}}\bigg)^{1/3}\bigg(\frac{I_L}{I_{14}}\cos\theta\bigg)^{2/3},
\end{equation}
where $n_c$ is the critical number density, $n_{351}$ is the critical number density for 351 nm light, and $I_{14}=10^{14}\textrm{W}/\textrm{cm}^2$.

While this may seem like a trivial modification to the Manheimer model, the result that the ablation pressure and mass ablation rate vary with different powers of $\cos\theta$ has important implications for controlling laser ablation when rays from multiple incidence angles are utilized. This can be seen easily by considering the mass ablation and acceleration of the two segments of the plane,  i.e. segment 1 and segment 2, illuminated by rays 1 and 2, depicted in Fig.~\ref{pddplane}B. 

The equations for mass ablation and acceleration of each of the planar segments are 
\begin{equation}
\frac{dM}{dt}=-L^2\dot{m}_a
\end{equation}
and
\begin{equation}
\frac{dv}{dt}=-\frac{P_aL^2}{M},
\end{equation}
where $L^2$ is the illuminated area. The solutions to these equations are 
\begin{align}\label{M}
&\frac{M(t)}{L^2}=\frac{M(0)}{L^2}\nonumber\\
&-0.61\frac{\textrm{g}}{\textrm{cm}^2}\bigg(\frac{A}{2Z}\frac{n_c}{n_{351}}\bigg)^{2/3}\cos^{4/3}\theta \ 
\end{align}
and
\begin{equation}\label{V}
v(t)=40.5\frac{\textrm{cm}}{\mu\textrm{s}}\bigg(\frac{I_L}{I_{14}}\frac{2Z}{A}\frac{n_{351}}{n_{c}}\bigg)^{1/3}\frac{1}{\cos^{2/3}\theta}\ln\bigg(\frac{M(t)}{M(0)}\bigg),
\end{equation}
where $M(0)/L^2$ is the t=0 areal mass of the segment. 
These equations can be used to determine a prescription for the laser intensity needed to drive the two adjacent planar regions of the plate with specific constraints. Several constraints can be specified, however, the three considered in this subsection are: (i) The masses of segments 1 and 2 are equal, i.e. $M_1(t)=M_2(t)$. (ii) The velocities of segments 1 and 2 are equal ($v_1(t)=v_2(t)$). (iii) The dynamic pressures of both segments are equal, assuming constant density, such that $M_1(t)v_1^2(t)=M_2(t)v_2^2(t)$. 

\begin{figure}
\includegraphics[scale=.43]{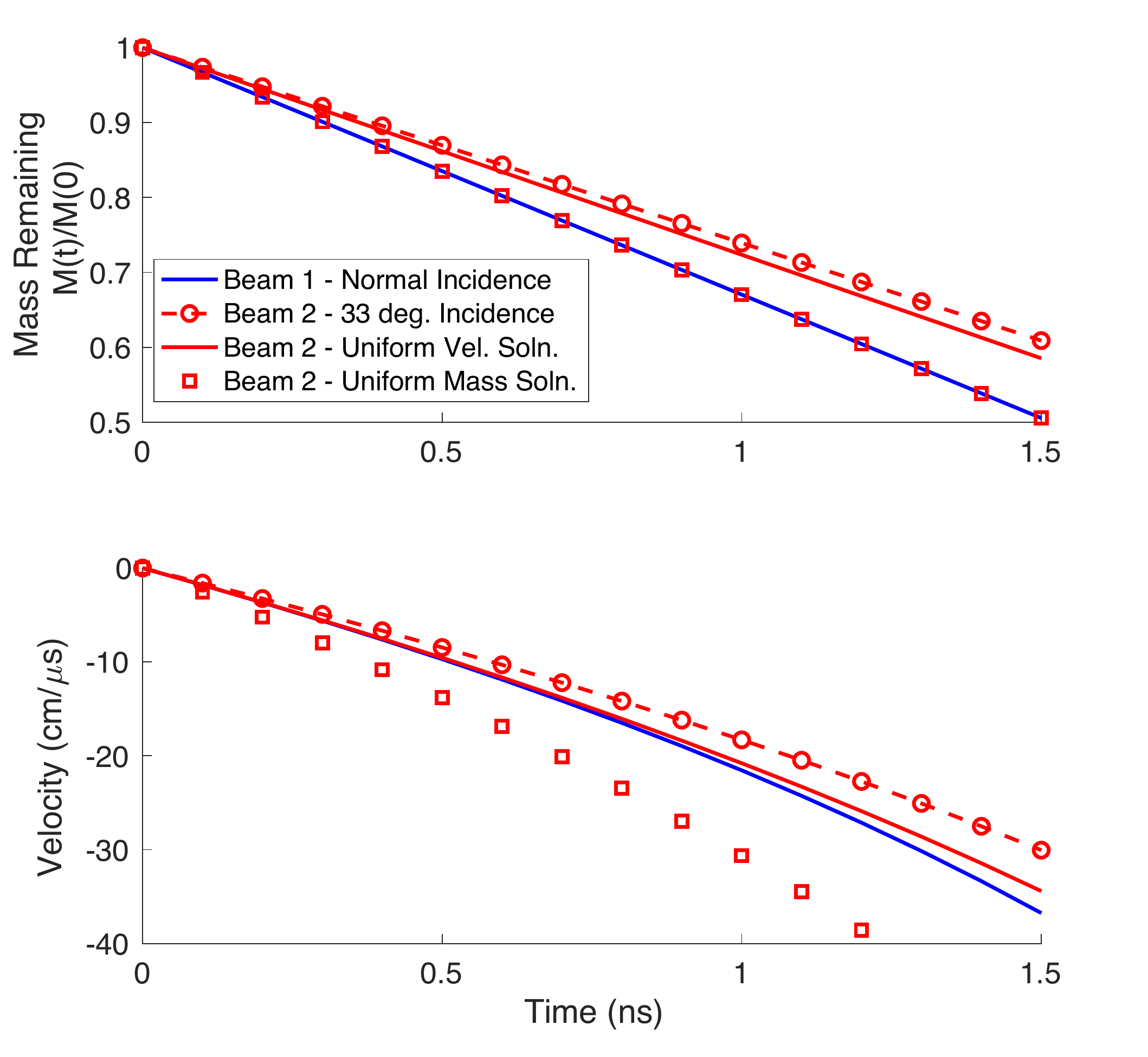}
\caption{Solutions for the mass remaining from Eq.~\ref{M} (top) and velocity from Eq.~\ref{V} (bottom) for different solutions in the text. \label{solnfig}}
\end{figure}

First consider case (i) of equal masses. Setting the solutions for the time dependent mass of each segment equal results in the time-independent result 
\begin{equation}\label{em}
I_2=\frac{I_1}{\cos^4\theta_2},
\end{equation}
where $\theta_2$ is the incidence angle of ray 2. 
Solutions of Eqs.~\ref{M} and \ref{V} with condition Eq.~\ref{em} are shown in Fig.~\ref{solnfig} for segments which are composed of a fully ionized, 15$\mu$m thick by 1 cm by 1cm beryllium plate, with $I_1=2\times10^{14}\textrm{W}/\textrm{cm}^2$, and $\theta_2=33^o$. The solutions (solid blue line for segment 1 and red squares for segment 2) have equal masses as expected, although the velocities of the two regions differ by more than 25 percent after 1.5 ns. A solution for $I_1=I_2$ is also shown (blue line for segment 1 and red circles with dashed line for segment 2).

In case (ii) the same procedure can be followed for equal velocities. By requiring equality of the velocities of each segment, it is apparent from Eq. \ref{V} that the required value of $I_2$ is a function of not only $\theta$ and $I_1$ but also of t. Nevertheless, an approximate solution for this situation can be obtained in the limit that the fractional mass remaining is close to unity. Expanding the logarithm in Eq. \ref{V} around $M(t)/M(0)=1$, keeping only the first two terms, results in the time independent solution
\begin{equation}\label{IV}
I_2\approx\frac{I_1}{\cos\theta_2}.
\end{equation}
Figure~\ref{solnfig} shows a solution for segment 2 using the intensity specification of Eq. \ref{IV} (red line). In the first 1.5 ns, the velocities match one another within a few percent while the difference in mass is $\sim$10 percent. This solution provides a significant improvement over that where $I_2=I_1$. 

It is important to note that the incidence angle scalings of the intensities given by Eq.\ref{em} and \ref{IV} are not due to geometrical $(\cos\theta)^{-1}$ reduction in intensity from the projection of the beam spot of an obliquely incident beam. Such effects do not occur in our 1D single ray model, but would need to be included in a spherical laser drive model. However, the effects of projection could be included in $I_1$ and $I_2$, but {\emph{here it is assumed that $I_1$ and $I_2$ are the surface intensities on the target due to rays 1 and 2}}. 

While a solution in which the velocity of the two regions may provide the most spatially uniform motion of the plate, a solution with equal dynamic pressure is also of interest. During deceleration of an imploding spherical shell by compressed fuel, the dynamic pressure of the shell is converted to internal pressure of the fuel, hence minimizing the variations in dynamic pressure improves the uniformity of the hot spot. Therefore, driving the spherical shell with dynamic pressure and velocity as uniformly as possible is desirable for obtaining better symmetry when the shell stagnates. Case (iii) evaluates the requirements for equal dynamic pressure in the planar system. The solutions for each segment given by Eqs.~\ref{M} and \ref{V} reveal that there is no time independent relation between $I_2$ and $I_1$ that will provide equal dynamic pressure across the plate unless both rays are normally incident. 

\begin{figure}
\includegraphics[scale=.4]{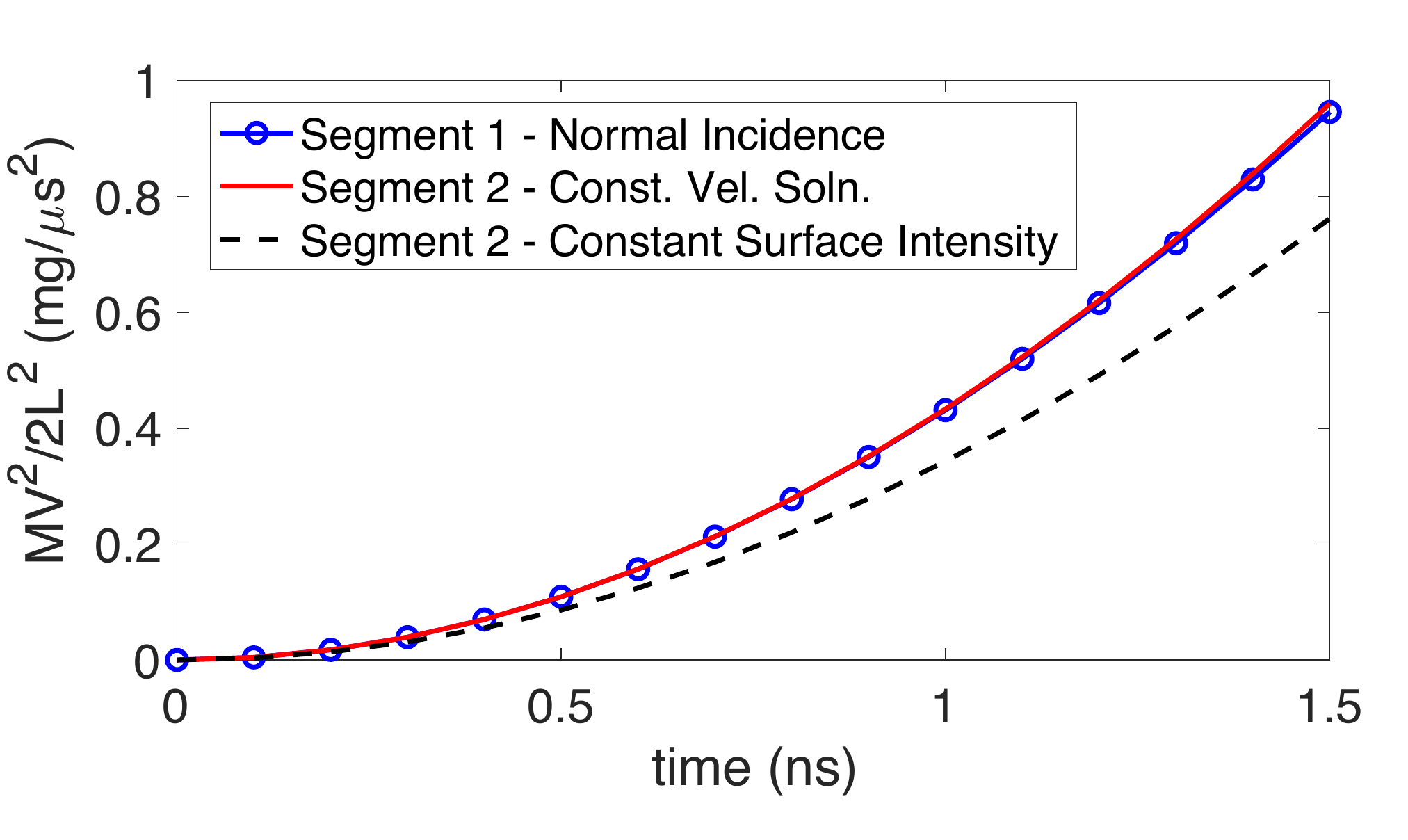}
\caption{Calculation of the areal dynamic pressure of two segments being driven with the intensity specified by Eq.~\ref{IV} for equal velocity in the limit of large mass remaining. The red and blue lines correspond to the mass fraction remaining and velocity indicated by the red and blue lines in each panel of Fig.~\ref{solnfig}.\label{dynpfig}}
\end{figure}

Although exact equality between dynamic pressure of the two segments is not possible, under certain circumstances approximate equality is possible. In cases where the ablated mass is small, the difference in dynamic pressure will be dominated by the velocity due to its quadratic scaling.  In this limit Eq.~\ref{IV} holds and an approximately equal dynamic pressure will be provided by this solution. Figure~\ref{dynpfig} shows the areal dynamic pressure for the solution given by Eq.~\ref{IV}, corresponding to the solutions shown in Fig.~\ref{solnfig}. Figure~\ref{solnfig} shows that the equal velocity solution is approximately valid for the first 1.5 ns. Correspondingly, the areal dynamic pressures in the two segments differ by approximately  1 percent at 1.5 ns. For comparison, a case with equal surface intensity to the normal incidence case on segment 2 is shown, which demonstrates that uniform surface intensity is an insufficient condition to drive a target uniformly within the model and that a surface intensity scaling of $1/\cos(\theta)$ is needed mitigate velocity variations for a given ray angle $\theta$. In addition, targets with smaller ablated mass fraction can have less variation in dynamic pressure imparted during early time when the Manheimer model is valid. 

\subsection{Conduction zone length and coronal temperature \label{czlen}}
The incidence angle modification to the flux balance of Eq.~\ref{B} not only changes the ablation pressure and mass ablation rate, but also the coronal temperature and the width of the conduction zone. The temperature resulting directly from Eq.~\ref{B} is 
\begin{equation}\label{cs}
\Gamma T_{tp}=c_s^2=\bigg(\frac{I_L}{4\rho_c\cos^2\theta}\bigg)^{2/3},
\end{equation}
which shows that the temperature increases with the -4/3 power of the incidence angle cosine. The larger temperature can be interpreted as a result of the same power being input into a lower density plasma. On NIF, 50 degree beams crossing the equator would have $\theta=40$ degrees resulting in a 43 percent increase in coronal temperature for the same surface intensity. Obviously, such estimates neglect the effects of CBET. Likewise, the modified version of Eq.~\ref{B} results in the width of the conduction zone being modified to
\begin{equation}\label{xaft}
x_{cz}=\frac{x_{cz0}}{\cos^{14/3}\theta},
\end{equation}
where $x_{cz0}$ is the conduction zone width at normal incidence given by Eq.~\ref{xaf}.

\section{Discussion}

\subsection{Implications for drive symmetry and symmetry evaluation\label{impl}}

In the example of Sec.~\ref{roleince}, it was shown that two segments of a plane, each driven by rays at different incidence angles, could not be driven with strictly equal velocity or dynamic pressure between regions without specifying a time dependent relation between the intensities of each ray.
Furthermore, without having all rays intercept the plane at the same incidence angle it was not possible to simultaneously provide equal dynamic pressure and velocity for each region even when utilizing a time dependent relation between $I_1$ and $I_2$. In this subsection, the implications for spherical PDD targets are explored, making the conjecture that ablation pressure and mass ablation rate will scale with different powers of $\cos\theta$ in the spherical case. Such an expectation is based on the analysis of the large spherical targets such as those shown in simulations of Sec.~\ref{val}. These results demonstrate that planar ablation scaling can be recovered when the length of the rarefaction is small compared to the target radius early in the ablation, provided that the laser deposition is localized. Given the arguments in Sec.~\ref{valinc}, it is reasonable to expect the same for oblique incidence.

\begin{figure}
\includegraphics[scale=.41]{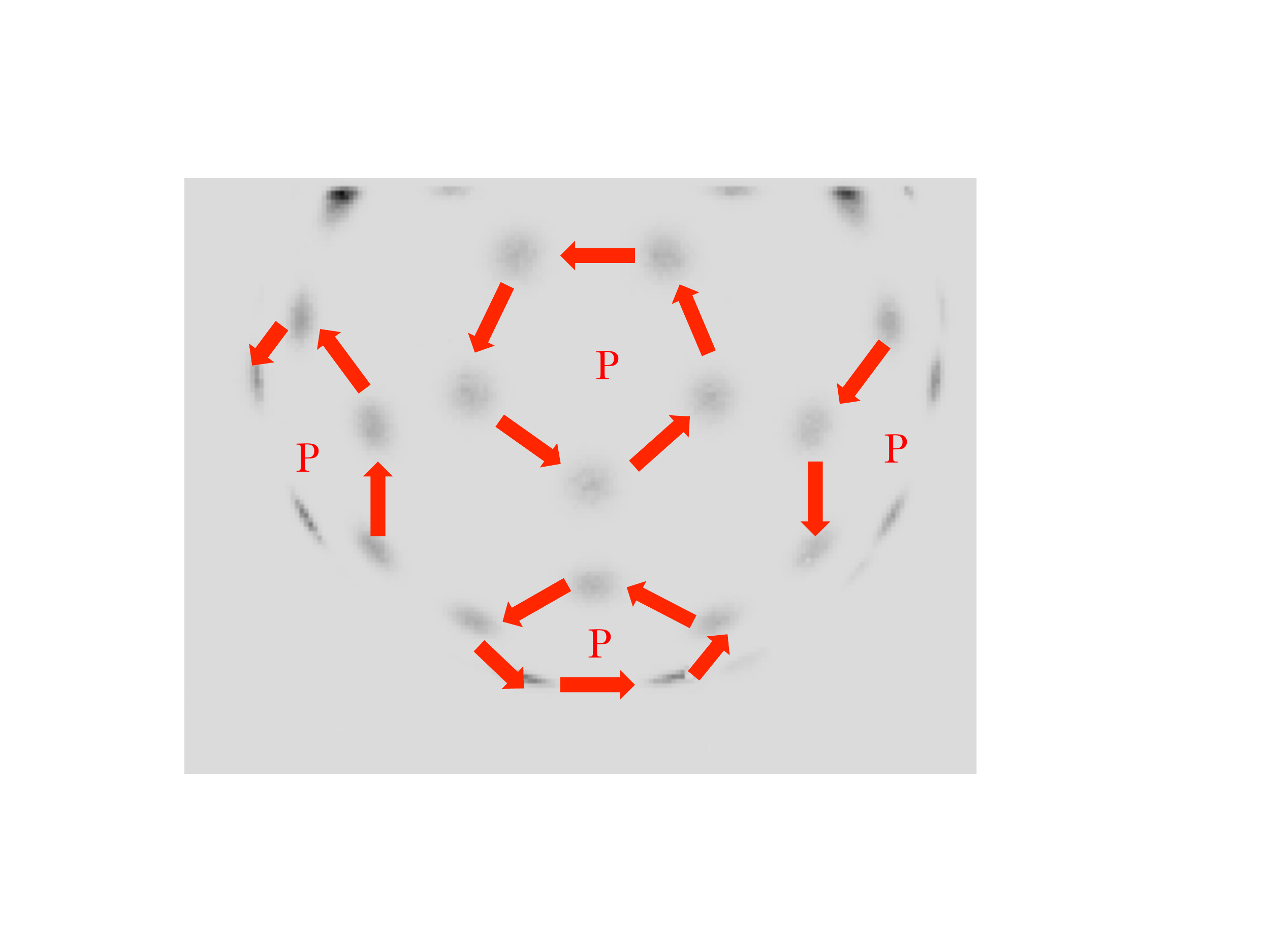}
\caption{ The clocked beam configuration on the 60 beam Omega Laser. Each beam is pointed to the projection of its neighboring beam port on the target surface, indicated by the red arrows. \label{clocked}}
\end{figure}

PDD targets are much more complex than the situation presented for the planar model, with several effects taking place simultaneously, obscuring the individual physics issues. Such effects may include, but are not limited to, the illumination by rays at several incidence angles within a single beam, different depths of absorption for beams overlapping on the target, isotropization of coronal temperatures due to electron conduction between illumination regions, refraction of rays, and laser plasma instabilities such as CBET. Such effects are difficult to model analytically and typically require treatment by radiation hydrodynamic simulations, especially in the case of the implosion of small spherical targets where the analogous equations of Sec.~\ref{ablmodel} are difficult to solve. Simulations and experiments may provide appropriate scalings for the mass ablation rate and ablation pressure of spherical targets driven by beams with centerlines at a single oblique incidence angle, even when the planar ablation limit is not recovered. One such beam configuration that may prove useful for such a study is the clocked beam configuration shown in Fig.~\ref{clocked}. The clocked beam configuration repoints each of the 60 beams on the Omega Laser to the projection of its counterclockwise neighboring beam port (within the same pent) on the sphere. Such configurations had been used in the first few years of PDD experiments to validate laser ray absorption at glancing incidence in simulations\cite{2004PhPl...11.2763S,2005PhPl...12e6304C}. However, detailed analysis of the underlying physics has not been presented.

If a significant difference in incidence angle scaling of the mass ablation rate and ablation pressure persist when these combined effects are considered, it would have important implications for the evaluation of PDD symmetry in present day experiments and for the design of future PDD schemes. For example, the degree of in-flight low Legendre mode ($\ell \lesssim10$) spatial uniformity is often used as a metric for implosion non-uniformity. However, a certain degree of uniformity in velocity does not imply the same degree of uniformity in dynamic pressure. The planar model presented in this paper provides a qualitative description of how the degree of uniformity in velocity and dynamic pressure can differ when beams at different incidence angles are present. 

The results of Sec.~\ref{roleince} also suggest that if the mass remaining is large, a drive scheme without a time dependent variation between beams may be possible. While this will require further work to determine, this would be fortuitous for the design of PDD pulse shapes at Omega, where the options for well controlled time dependent variation between beams is limited. In many cases, however, the amount of fractional mass remaining in the shell differs significantly from unity. When this is the case, the initial mass of the shell can be contoured with polar angle, tailoring a more uniform stagnation radius, albeit only for known very low mode variations. Although this will not guarantee better in flight symmetry, such methods have been recently demonstrated to reduce low mode asymmetry at stagnation in PDD experiments at Omega\cite{2016PhPl...23a2711M}.

\subsection{Incidence angle and imprint in experiment\label{impr}}

In Sec.~\ref{czlen}, the width of the conduction zone was shown to have a -14/3 power law scaling with the cosine of the incidence angle. As the width of the conduction zone increases so does the smoothing of the pressure applied to the shell, owing to conduction and fluid motion in the transverse direction to the target normal. As a result, the amplitude of the mass modulation due to laser imprint is expected to decrease with increasing incidence angle. The only study known to the authors that demonstrates any of the effects of incidence angle presented in this paper is that of Smalyuk et al.\cite{2005PhPl...12d0702S}. Therefore, this subsection is dedicated to a qualitative comparison of their experimental measurements with the predictions for the conduction zone in this paper.    

In their experiment, a planar CH target was illuminated to $1\times10^{14} W/cm^2$ using 6 beams of the Omega laser. Five of the beams (drive beams) utilized standard distributed phase plates (DPPs), while one probe beam utilized a special DPP that produced a spatial sinusoidal $60\mu m$ wavelength intensity variation in one direction. The laser pulse of the drive beams was offset to lag the probe beam by up to 220 ps, i.e. starting at t=-220ps, allowing the probe beam to imprint the sinusoidal modulation onto the target. Since the initial modulation on the target is too small to be detected, a second $60 \mu m$ wavelength modulation of know amplitude was machined into the target, perpendicular to the imprinted laser intensity modulation. As the drive beams turned on, both modulations grew linearly due to Rayleigh-Taylor growth at the unstable ablation surface. Once at large enough amplitude, the machined and laser imprinted modulation were measured via backlighter and were observable due to their modulation of the material's optical depth. Hence, the initial modulations from the probe beam could be calculated from the drive-beam-enhanced modulations using
by $\eta_{las}(k,t=0)=\eta_{mac}(k,t=0)\delta OD_{las}(k,t)/\delta OD_{mac}(k,t)$, where $\eta_{las}$ and $\eta_{mac}$ are the modulation amplitudes imprinted by the laser at the beginning of the drive beam pulse at $t=0$ and those pre-machined into the target, respectively. The corresponding optical depth modulations are denoted by $\delta OD(k,t)$. The imprint efficiency (defined as the imprint modulation amplitude normalized to the relative intensity variation of the probe beam $E(k)=\eta_{las}(k,t=0)/(\delta I(k)/I)$) was measured for three different probe beam incidence angles of $23^o$, $48^o$, and $58^o$, with a factor of 3 decrease in imprint efficiency measured over the incidence range. In their paper, they concluded from their radiation hydrodynamic simulation analysis that the decrease in imprint resulted from a greater distance between the laser absorption region and the ablation surface. The measured decrease in imprint efficiency and the increase in distance between the ablation surface and laser absorption region are both qualitatively consistent with the analytical prediction of a widening of the conduction zone with incidence angle in Eq.~\ref{xaft}.

\section{Conclusion\label{conc}}

This paper demonstrates the effect of laser incidence angle on the ablation pressure and mass ablation rate in the planar Manheimer model of laser ablation. For oblique incidence, calculations demonstrated that laser energy deposition occurs closer to the ray turning point than in the normal incidence case, suggesting that the local energy deposition is appropriate in the first few nanoseconds of the laser pulse. With the effect of incidence angle included, it was found that the ablation pressure and mass ablation rate scale with the 4/3 and 2/3 power of cosine of the incidence angle. A simplified segmented planar system driven by rays at different angles was used to demonstrate the need for varying intensities, and in some cases time dependent intensity profiles, to achieve equivalent velocities of the two segments. Such considerations were needed to provide equal dynamic pressure in the model as well. 

In addition to the ablation pressure and mass ablation rate, an analytic prediction for the width of the conduction zone as a function of incidence angle was provided. The prediction of the widening of the conduction zone with increasing incidence angle is consistent with prior experimental measurements of imprint efficiency at oblique incidence. 

While only the planar system was studied in detail here, the implications for spherical targets were discussed assuming that similar scalings of ablation pressure and mass ablation rate hold. Such assumptions are not unfounded. The evaluation of 1-D simulations of the normal ablation of a spherical 3.1 mm radius target indicate that planar Manheimer ablation is recovered in the first $\sim1-2$ ns of the laser pulse when the target is most susceptible to imprint from drive variations. In these situations, accounting for differences in drive from oblique beams may decrease the amplitude of spherical harmonic modes synergistic with the number of cone angles. Further work will be required to determine the exact incidence angle scaling of ablation pressure, mass ablation rate, and conduction zone length in the ablation of spherical targets when multiple effects such as beams with centerlines at multiple angles, ray refraction, coronal heat conduction, and CBET are included.

\section*{Acknowledgments}
The authors thank the referee for their detailed comments and suggestions. This work was supported by the Laboratory Directed Research and Development program of Los Alamos National Laboratory under project number 20180051DR.

\appendix*
\section{Absorption Fraction}

The absorption fraction for an electromagnetic wave initially propagating at incidence angle $\theta$ in the isothermal rarefaction density profile given by Eq.~\ref{ssrho} is
\begin{equation}\label{al0}
A_L=1-\exp\big(-\frac{8}{3}\frac{\nu_{ec}}{c}L\cos ^3\theta\big),
\end{equation}
This is the $x=0$ result of Eq.~\ref{alx}. Equation \ref{al0} is a well known formula, variations of which can be found in textbooks\cite{kruer1988physics,atzeni2004physics} and the literature\cite{1982PhFl...25.1051M}. The absorbed light fraction can be determined as a function of incidence angle, intensity $I$, and length scale L can be determined by using the power balance relation of Eq.~\ref{B} with the absorbed light fraction 
\begin{equation}
A_LI_L=4\rho_c\cos^2\theta (\Gamma T_e)^{3/2}
\end{equation}
and inserting the obtained temperature relation into 
\begin{equation}\label{nuec}
\frac{\nu_{ec}}{c}\approx\frac{25 Z}{\lambda_L^2T_e^{3/2}}\textrm{cm}^{-1}, 
\end{equation}
where Z is the ionization state and $\lambda_L$ is the laser wavelength in $\mu$m\cite{atzeni2004physics}. The resultant relation is
\begin{equation}\label{Al}
A_L=1-\exp\big(-\frac{I^*_L}{A_LI_L}\big)
\end{equation}
where
\begin{equation}
I^*_L=1.4\times10^{11}\frac{ZL}{\lambda_L^4}\cos^5\theta  \ \ \ \textrm{W}/\textrm{cm}^2
\end{equation}
is a reference intensity. Equation \ref{Al} can be solved numerically for $A_L$. This method was originally used by Mora\cite{1982PhFl...25.1051M} for the case of normal incidence. Typically when $I^*_L/I_L\gg1$ the absorption is described by strong inverse bremsstrahlung, while when $I^*_L/I_L\ll1$ the absorption is described by the Manheimer model\cite{1982PhFl...25.1051M}. The scaling of Manheimer ablation can still be exhibited with an intensity at the transition between Manheimer and SIB ablation if the laser deposition is strongly peaked at the laser turning point. 

\begin{figure}[h!]
\includegraphics[scale=.43]{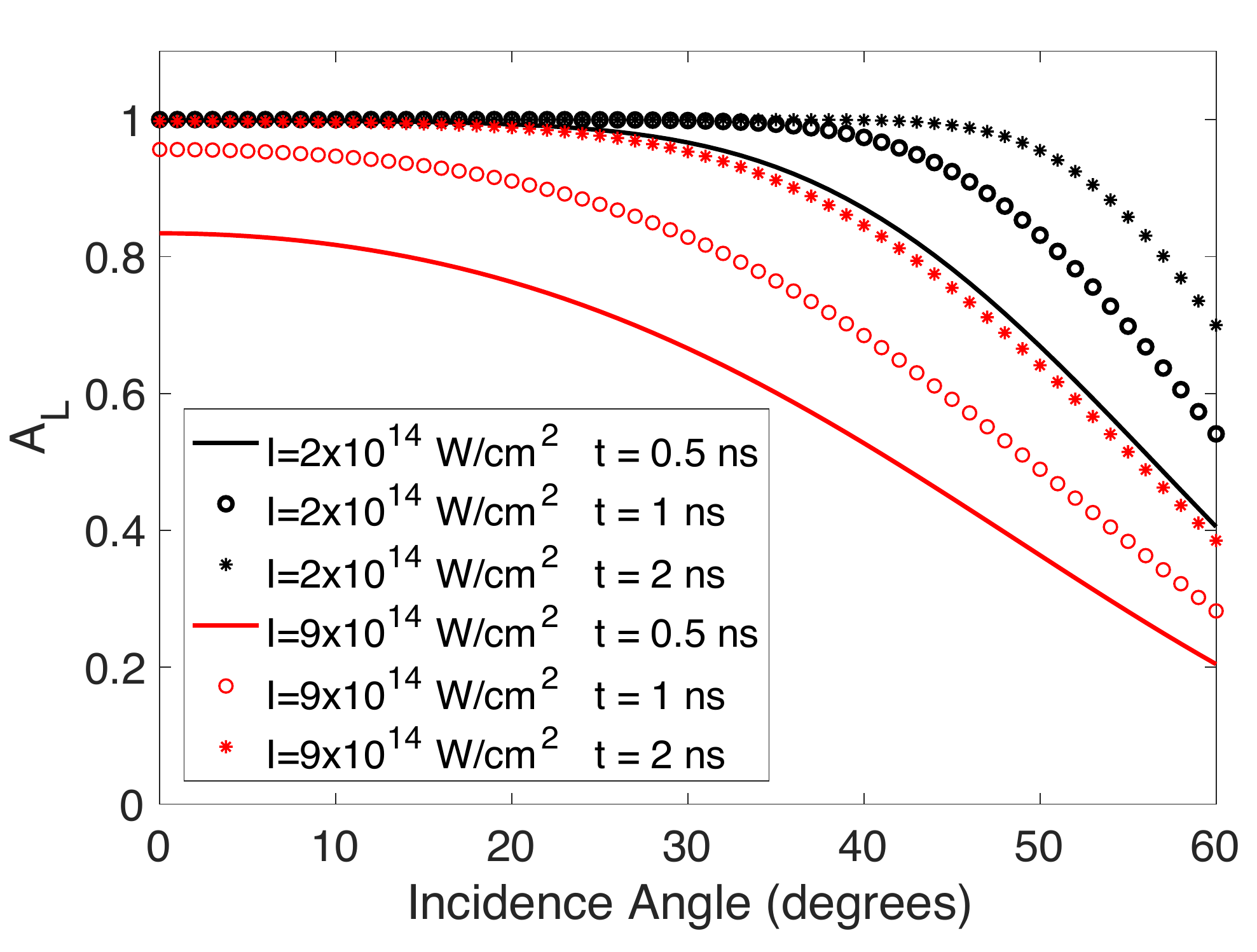}
\caption{The calculated absorbed light fraction as a function of incidence angle at t=0.5, 1, and 2 ns for intensities of $2\times10^{14}$ and $9\times10^{14}$ $\textrm{W} \ \textrm{cm}^{-2}$. \label{figal}}
\end{figure}

In this paper, the example considered was at an intensity of $2.65\times10^{14}\ \ \textrm{W}/\textrm{cm}^2$, however, intensities up to $\sim10^{15} \ \ \textrm{W}/\textrm{cm}^2$ are often encountered in direct drive ignition designs. To cover the range of intensities encountered, the absorbed light fraction is calculated for intensities of  $2\times10^{14}\ \ \textrm{W}/\textrm{cm}^2$ and $9\times10^{14}\ \ \textrm{W}/\textrm{cm}^2$ assuming a fully ionized Be plasma with a electron temperature of 2keV, and a 351 nm laser radiation wavelength. With such considerations, the length scale at 1ns is $L=73 \ \mu m$ and the reference intensity at the corresponding time is $I_L^*=2.70\times10^{15}\cos^5\theta  \ \textrm{W}/\textrm{cm}^2$. The calculated absorption coefficients as a function of incidence angle at 0.5 ns, 1 ns, and 2 ns are shown for both intensities in Fig.~\ref{figal}. For the intensities presented, there is little change in $A_L$ over the first 30 degrees. In this paper it will be assumed that the absorbed energy fraction varies little with incidence angle. This assumption is valid for low intensity such as the example considered in Sec.~\ref{drive}, however, the absorbed light fraction should be included for higher intensity.

\bibliography{incidence}
\end{document}